\documentclass[final,3p,times]{elsarticle}
\usepackage{graphicx,color}
\usepackage{amssymb,amsmath,array}
\biboptions{sort,compress}

\journal{Physics Letters B}

\usepackage{epsfig}
\usepackage{mathptmx}      
\usepackage{latexsym}

\def\pslash{p\kern-.45em\slash}
\def\qslash{q\kern-.35em\slash}

\def\1{\'{\i}}

\def\XXint#1#2#3{{\setbox0=\hbox{$#1{#2#3}{\int}$}
     \vcenter{\hbox{$#2#3$}}\kern-.5\wd0}}

\newcommand{\simge}{\hspace*{0.2em}\raisebox{0.5ex}{$>$}
     \hspace{-0.8em}\raisebox{-0.3em}{$\sim$}\hspace*{0.2em}}

\begin{document}

\begin{frontmatter}

\title{Two-body double pole and three-body bound states: 
\\ 
physical and unphysical quark masses}

\author[1]{V.S. Tim\'oteo\footnote{Corresponding author: 
varese@unicamp.br}}

\author[2,3]{U. van Kolck}
 
\address[1]{Grupo de \'Optica e Modelagem Num\'erica - GOMNI, Faculdade de Tecnologia - FT, \\
Universidade Estadual de Campinas - UNICAMP, 13484-332, Limeira, SP, Brazil}

\address[2]{Universit\'e Paris-Saclay, CNRS/IN2P3, IJCLab, 91405 Orsay, France}

\address[3]{Department of Physics, University of Arizona, Tucson, Arizona 85721, USA}

\begin{abstract}

We solve the Faddeev bound-state equations for three particles with simple two-body nonlocal, separable potentials that yield a scattering length twice as large as a positive effective range, as indicated by some lattice QCD simulations. Neglecting shape parameters, the two-body bound state is a double pole. For bosons we obtain a correlation between three- and two-body energies. 
For nucleons, this correlation depends additionally on the ratio of effective ranges in the two two-body $S$-wave channels. 
When this ratio takes the value suggested by lattice QCD, our three-body energy agrees well with a direct lattice determination. 
When this ratio takes the experimental value, we find a three-body bound state with energy close to that of the physical triton. 
We suggest that results could be improved systematically with distorted-wave perturbation theory around a separable potential whose form factor is an inverse square root of momentum squared.
\end{abstract}

\end{frontmatter}

\section{Introduction}

Great progress has been achieved in the simulation of light nuclear systems with lattice Quantum Chromodynamics (LQCD) since the first fully dynamical calculation by the NPLQCD collaboration appeared \cite{Beane:2006mx}. A seminal achievement \cite{NPLQCD:2012mex} was the calculation of the binding energies for $A\leq 4$ nucleons in the limit of SU(3) symmetry at the physical strange quark mass, corresponding to an unphysical pion mass $m_\pi\simeq 806$ MeV. The relatively shallow $A\leq 3$ bound states could be described at leading order (LO) of an effective field theory (EFT) without explicit pion fields (Pionless EFT) patterned on the real world \cite{Hammer:2019poc}. The EFT postdicted an $A=4$ state consistent with the LQCD result \cite{Barnea:2013uqa,Kirscher:2015yda,Contessi:2017rww,Bansal:2017pwn} 
and predicted binding energies for $A=5,6,16,40$ nuclei \cite{Barnea:2013uqa,Contessi:2017rww,Bansal:2017pwn} as well as $A=3$ scattering parameters \cite{Kirscher:2015yda}. This example illustrates how one day we might be able to use LQCD to determine nuclear EFT parameters and then derive nuclear properties from the solution for the long-range EFT dynamics, without input from low-energy experimental data.

At the same quark masses, the $A=2$ effective-range-expansion (ERE) parameters have also been obtained \cite{NPLQCD:2013bqy,Wagman:2017tmp} using L\"uscher's method \cite{Luscher:1986pf,Luscher:1990ux}.
The remarkable result was that in both two-nucleon $S$-wave channels ($^1S_0$ and $^3S_1$) the scattering length turned out to be, within one or two standard deviations, twice as large as the {\it positive} effective range, $a_2/r_2=2$, while shape parameters were found small. (Considering errors,
in strange channels the ratio is not very far from 2 either \cite{Wagman:2017tmp}.) As noted in Ref. \cite{Barnea:2013uqa}, these parameters imply an unusual double pole at positive imaginary (relative) momentum. A subsequent analysis \cite{Berkowitz:2015eaa} gave instead $a_2/r_2\approx 3$, amounting to two single poles at positive imaginary momentum, one of which is a shallow bound state, the other a deeper ``redundant'' pole \cite{Ma:1946,terHaar:1946}, as noted in Ref. \cite{Iritani:2017rlk}. In contrast, Ref. \cite{Horz:2020zvv} finds $a_2/r_2\approx -1$, with the shallower pole now a virtual state. Interpolating-operator dependence is significant in these calculations \cite{Amarasinghe:2021lqa} and there is no consensus in the LQCD community about the proper pole configuration of the two-nucleon system at unphysical pion masses --- for a recent discussion and references, see Ref. \cite{Nicholson:2021zwi}. 

A common characteristic of these LQCD calculations, which might survive improvements, is a positive effective range that is large on the scale of $m_\pi^{-1}$, which presumably determines the range of the interaction. It is not difficult to incorporate a large and {\it negative} effective range in Pionless EFT, which then excludes redundant poles \cite{Habashi:2020qgw,vanKolck:2022lqz}. As we discuss in this paper, allowing for a large and positive effective range with negligible shape parameters --- and thus two shallow two-body poles --- demands a reformulation of Pionless EFT: a resummation of derivative interactions and thus a nonlocal potential. In this case, only one $A=2$ bound state exists; the other pole (if not degenerate) is redundant in the sense of not being associated with a normalizable wavefunction.

When two shallow $S$-wave poles coalesce into a double pole on the positive imaginary axis of the complex-momentum plane, the theory at LO has a single parameter, which we can take to be the effective range, or alternatively the two-body binding energy $B_2$. Contrary to the standard pionless power counting, the new LO potential ensures many-body systems have well-defined ground states without a three-body force. The LO ratios $2B_A/AB_2 \equiv \varsigma_A$ are then universal numbers, independent of the details of the short-range potential. Here we compute $\varsigma_3$ in the case of a single $S$-wave two-body channel. 

When two $S$-wave channels (labeled 0 and 1) are present, as in the two-nucleon system, the $\varsigma_A$ are functions only of the ratio of two-body binding energies, $B_{21}/B_{20}$. 
For nucleons, we entertain the possibility that the pole structure is in some sense close to the double-pole situation found in Refs. \cite{NPLQCD:2013bqy,Wagman:2017tmp} and ask what its implications are for the three-nucleon system from an EFT perspective. 
At unphysical pion mass, we predict the three-nucleon binding energy and compare it with the result from Ref. \cite{NPLQCD:2012mex}.

The same LO potential, with a different constraint among two-body scattering lengths and effective ranges, has been arrived at by quite different considerations in Ref. \cite{Beane:2021dab} and applied to the two-nucleon system at physical pion mass to next-to-leading order (NLO). 
Similar approaches for nuclear systems with explicit pion fields \cite{SanchezSanchez:2020kbx,Yang:2020pgi,Yang:2021vxa,Peng:2021pvo}
and for halo nuclei with $P$-wave nucleon-core interactions \cite{Li:2023hwe} have also been explored recently.
We therefore consider here also the triton binding energy in the real world. At physical pion mass, $a_2^{(^3S_1)}/r_2^{(^3S_1)}\simeq 3.2$ and $a_2^{(^1S_0)}/r_2^{(^1S_0)}\simeq -8.8$. These numbers are not particularly close to the value of $2$ corresponding to the double pole. However, the relatively large scattering lengths are due to fine tuning and hopefully can be corrected at higher orders.

This letter is organized as follows. In Sec. \ref{2bsys} we present the two-nucleon properties of our separable potential in LO; we also introduce two other separable potentials that we use subsequently to gauge the sensitivity to higher ERE parameters. Results for the system of three spinless particles are presented in Sec. \ref{3spinless}. The generalization to three nucleons in the triton channel is given in Sec. \ref{3nucleons}. It is applied to unphysical $m_\pi\simeq 806$ MeV and physical $m_\pi\simeq 140$ MeV in Secs. \ref{unphysical} and \ref{physical}, respectively. Conclusions can be found in Sec. \ref{conc}.

\section{Two-body system}
\label{2bsys}

We are concerned with two particles of mass $m\gg R^{-1}$ which form $S$-wave bound states with size $\aleph^{-1}\gg R$, where $R$ is the range of the underlying interaction. The standard power counting of Pionless EFT \cite{Hammer:2019poc} is designed to account for a single $T$-matrix pole, which can be such a shallow bound or virtual state. It demands a single, nonderivative contact interaction at LO, which scales as $\aleph^{-1}$ \cite{vanKolck:1997ut,Kaplan:1998tg,Kaplan:1998we,vanKolck:1998bw}. Two shallow $S$-wave poles can be accounted for with an additional two-derivative interaction at LO, which scales as $\aleph^{-3}$ \cite{Habashi:2020qgw,vanKolck:2022lqz}. Renormalization requires $r_2\le 0$ \cite{Phillips:1997xu,Beane:1997pk} and, in the various possible pole configurations that result, at most one pole is on the positive imaginary momentum axis and corresponds to a bound state \cite{Habashi:2020qgw,vanKolck:2022lqz}. In both cases, the EFT is local in the sense that at any order only a finite number of derivatives enter interactions at any given order. 

Unless we allow for energy-dependent interactions, two shallow $S$-wave poles on the positive imaginary axis demand LO nonlocality in the sense of an infinite number of contact interactions. The origin of this strong nonlocality is nebulous at best. It appears to go against the principles of EFT where the only nonlocal forces are produced by the exchange of virtual particles, which are kept as explicit degrees of freedom. Assuming nevertheless that this situation can be realized in physical systems, all contact interactions must have large, correlated parts whose strengths are set by powers of $\aleph^{-1}$, so that they should be resummed into a nonanalytic function. It is simplest to search for a combination of contact interactions that produces an (energy-independent) separable two-body potential supporting only these two poles. In terms of the magnitudes of the incoming and outgoing relative momenta $\vec{p}$ and $\vec{p}'$, this type of two-body potential is written as \cite{Yamaguchi:1954mp}
\begin{equation}
V_2(p',p) = \frac{4\pi}{m}\,\lambda\, g(p')\, g(p)
\end{equation}
with a strength $\lambda$ and a real function $g(p)$ obeying $g(0)=1$. The two-body $T$ matrix at energy $E=k^2/m$ is then
\begin{equation}
T_2(p',p;k) 
=\frac{4\pi}{m} \frac{g(p')g(p)}{\Lambda^{-1}(-ik)-\lambda^{-1}}
= \frac{4\pi}{m}\frac{g(p')g(p)}{g^2(k)}
\left[-ik -\frac{1}{a_2}+ R(k)\, k^2\right]^{-1}, 
\label{T2}
\end{equation}
where
\begin{equation}
\frac{1}{a_2}=\frac{1}{\lambda} +\frac{2}{\pi} \int_0^\infty dl \,g^2(l),
\quad
R(k)=\frac{1}{a_2 k^2} \left(g^{-2}(k)-1\right)
+\frac{i}{k}-\frac{2}{\pi}g^{-2}(k)
\int_0^\infty dl\,\frac{g^2(l)}{l^2-k^2-i\varepsilon},
\label{EREpar}
\end{equation}
and
\begin{equation}
\Lambda^{-1}(-ik)=-\frac{2}{\pi} \int_0^\infty dl \,g^2(l) 
-ik +R(k) k^2.
\end{equation}
Some of the poles of Eq. \eqref{T2} are determined by the form of $g(p)$ alone, while others depend on the strength $\lambda$. The strength that produces a pole on the positive imaginary axis, $k=i\kappa_2$ with $\kappa_2>0$, satisfies
\begin{equation}
\lambda = \Lambda(\kappa_2).
\label{bs2}
\end{equation}
This relation can also be obtained from the Schr\"odinger equation, which yields a momentum-space wavefunction \cite{Yamaguchi:1954mp}
\begin{equation}
\psi_2(p) =N \frac{g(p)}{p^2+\kappa_2^2}, 
\quad
N= \left[\frac{1}{2\pi^2} 
\int_0^\infty dp \, \frac{p^2  g^2(q)}{(p^2+\kappa_2^2)^2}\right]^{-1/2}.
\label{sepwf}
\end{equation}

A single shallow $S$-wave pole arises at $\kappa_2= 1/a_2$ for 
$\lambda={\cal O}(\aleph^{-1})$. Regularization can be effected with a momentum cutoff $\Lambda$ introduced at intermediate stages of the calculation; one way to do so is through a $g(p)$ which obeys $g(p\gg \Lambda)\to 0$ and is viewed as just a regulator. After renormalization, when the $\Lambda$ dependence of $\lambda$ is chosen appropriately --- for example, so that $\kappa_2$ in Eq. \eqref{bs2} be finite --- the on-shell $T$ matrix $T_2(k,k;k)$ takes the form of the ERE truncated at the scattering length, $R(k)$ being arbitrarily small. To generate instead the ERE truncated at the effective range, which leads to two poles, we can take \cite{Gourdin:1957,Bander:1964gb,Beane:1997pk}
\begin{equation} 
g(p)=\left(1+\frac{p^2}{\alpha^2}\right)^{-1/2},
\label{sqrtY}
\end{equation}
with a parameter $\alpha>0$, such that 
\begin{equation}
\frac{1}{a_2}=\frac{1}{\lambda} +\alpha,
\quad
2R(k) = r_2 = -\frac{2}{\lambda\alpha^2}.
\label{sqrtYfeatures}
\end{equation}
If $\alpha=\Lambda$, this $g(p)$ is but one example of a regulator in the EFT for a single pole. In contrast, if $\alpha={\cal O}(\aleph)$ is a physical parameter, the form factor \eqref{sqrtY} represents correlated parts of all higher-derivative contact interactions, which are now LO: the effective range is finite and $r_2>0$ for $\lambda<0$. On shell, the two-body $T$ matrix is the same as that obtained with a dimer field \cite{Kaplan:1996nv,Beane:2000fi}, which is a ghost for $r_2>0$. Off shell, it decreases faster with momenta on account of the form factor. 

This separable potential generates two poles on the positive imaginary momentum axis for $\lambda <-1/\alpha$: {\it i)} a pole at $\kappa_2=\alpha$, which is redundant in the sense of being independent of $\lambda$, and {\it ii)} a bound state at $\kappa_2$ related to $\lambda$ through
\begin{equation}
\Lambda^{-1}(\kappa_2) =-\frac{\alpha^2}{\kappa_2+\alpha}.
\label{bs2sqrtY}
\end{equation}
For $-2/\alpha<\lambda <-1/\alpha$, the bound state is the shallower pole, at which the $S$ matrix has a negative imaginary residue. For $\lambda <-2/\alpha$ the bound state is deeper than the redundant pole. The opposite sign for the $S$-matrix residue translates into an opposite sign for the full two-body propagator, which in turn means that if the state were considered as elementary it would require an imaginary coupling to the two particles \cite{Bander:1964gb}. 
\label{sqrtYwf}
The two possibilities arise from two choices in a one-parameter family of separable potentials that reproduce the ERE at the effective-range level \cite{Gourdin:1957}, of which the form \eqref{sqrtY} is the simplest. The situation here is the counterpart to the phase-equivalent, exponentially decreasing local potentials of Ref. \cite{Bargmann:1949}, where different choices of parameters can make either the shallower or the deeper pole a bound state. In-between these two cases, a double pole arises at 
$\kappa_2=\alpha=1/r_2=2/a_2$ for $\lambda=-2/\alpha$. Despite the sign of the $S$-matrix residue, the wavefunction
\begin{equation}
\psi_2(r) = 
4\left(\frac{\kappa_2}{\pi}\right)^{3/2} K_0(\kappa_2 r),
\label{sqrtYwfdoulepole}
\end{equation}
where $K_0(x)$ is the modified Bessel function of the second kind, 
decreases with distance albeit as $\exp(-\kappa_2 r)/\sqrt{r}$.

NLO two-body corrections, which allow for a perturbative shape parameter (the coefficient of the $k^4$ term in the ERE), have been discussed in Refs. \cite{Beane:2021dab,Peng:2021pvo}. For the purposes of benchmarking and of gauging the effects of higher ERE parameters in the three-body system, we consider here also two other separable potentials with form factors $g(x)$ that go to 0 increasingly faster at large $x$. One choice is the original Yamaguchi potential \cite{Yamaguchi:1954mp}, 
\begin{equation}
g(p)=\left(1+\frac{p^2}{\alpha^2}\right)^{-1},
\end{equation}
for which
\begin{equation}
\frac{1}{a_2}=\frac{1}{\lambda} + \frac{\alpha}{2},
\quad
r_2 = \frac{1}{\alpha} \left(1-\frac{4}{\lambda\alpha}\right),
\quad
2R(k)=r_2-\frac{2k^2}{\lambda \alpha^4}.
\label{Yamaguchifeatures}
\end{equation}
The Yamaguchi potential generates a (dimensionless) shape parameter 
$P_2
=-(\lambda \alpha)^{-1} (1-4/\lambda\alpha)^{-3} r_2^{3}$, 
and for $\lambda<-2/\alpha$ there are three poles on the positive imaginary momentum axis: a double pole at $\kappa_2=\alpha$ independently of $\lambda$, and a bound state related to $\lambda$ by
\begin{equation}
\Lambda^{-1}(\kappa_2) =-\frac{\alpha^3}{2(\kappa_2+\alpha)^2}.
\label{bs2sqrtY}
\end{equation}
This bound state has the usual $S$-matrix residue sign and a Hulth\'en-type wavefunction \cite{Yamaguchi:1954mp}.
It is shallower than the double pole as long as $\lambda>-8/\alpha$, and found at $\kappa_2=(\sqrt{2}-1)\alpha=2(\sqrt{2}-1)/r_2
\simeq 0.828/r_2$
for $\lambda=-4/\alpha$, when $a_2=2r_2$. 

The other choice is the toy potential of Refs. \cite{RuizArriola:2013wvi,RuizArriola:2014gam,RuizArriola:2014lyp,RuizArriola:2016zmo,Timoteo:2021ika},
which consists of a Gaussian form factor
\begin{equation}
g(p) =  e^{- p^2/\alpha^2},
\label{gau}
\end{equation}
for which
\begin{eqnarray}
&&\frac{1}{a_2}=\frac{1}{\lambda}+ \frac{\alpha}{\sqrt{2\pi}},
\quad
r_2=\frac{4}{\alpha}\left(\frac{1}{\sqrt{2\pi}} - \frac{1}{\lambda\alpha}\right),
\nonumber\\
&&2R(k)= r_2+ \frac{4}{\alpha}
\left[\frac{\alpha^2}{2k^2} \left(e^{2k^2/\alpha^2}-1-\frac{2k^2}{\alpha^2}\right)
+\sqrt{\frac{2}{\pi}}\left(\int_0^1 dt \, e^{2k^2t^2/\alpha^2} -1\right)\right],
\label{Gaussianfeatures}
\end{eqnarray}
and 
\begin{equation}
\Lambda^{-1}(\kappa_2)=- \frac{\alpha}{\sqrt{2\pi}}
+\kappa_2 \, \exp\left(2\kappa_2^2/\alpha^2\right)
\, \mathrm{erfc} \left(\sqrt{2}\kappa_2/\alpha\right),
\label{lambda2Gauss}
\end{equation}
where $\mathrm{erfc}$ is the complementary error function. The Gaussian potential gives rise to all ERE parameters and, since $g(p)$ has no poles, all poles of the $T$ matrix on the positive imaginary axis are solutions of Eq. \eqref{lambda2Gauss}, {\it i.e.} bound states. The attractive potential that generates $a_2=2r_2$ has $\lambda=- \sqrt{2\pi/(1-\pi/4)}/\alpha \simeq -5.41/\alpha$. For this value there is a bound state with $\kappa_2\simeq 0.795/r_2$.

While in general a choice of $a_2$ and $r_2$ leads to two possible set of values for $\alpha$ and $\lambda$, for $a_2=2r_2$ only one set produces finite-energy bound states. The constraint $a_2=2r_2$ imposes a relation between $\lambda$ and $\alpha$: 
$\lambda \alpha= -c$, with $c= 2,4,\sqrt{2\pi/(1-\pi/4)}$ for square-root, Yamaguchi, and Gaussian forms respectively. This leaves a single independent, dimensionful parameter, which we may choose as $r_2$. The ground-state binding energy can be written as
\begin{equation}
B_2= \frac{\beta_2}{mr_2^2},
\label{B2}
\end{equation}
where the dimensionless number $\beta_2$, which depends on the potential, is given in Table \ref{beta2table}. 

\begin{table}[tb]
\begin{center}
\begin{tabular}{c|c|c|c|c}
  &Square-Root & \text{Yamaguchi} & \text{Gaussian}    
  \\ 
  \hline
  $\beta_2$ & 1 &  $4 \left(\sqrt{2}-1\right)^2$& 0.63
\end{tabular}
\caption{
\label{beta2table}
Values for the coefficient $\beta_2$ of the correlation \eqref{B2} between the two-body binding energy and the effective range for the square-root, Yamaguchi, and Gaussian separable potentials, when $a_2=2r_2$.
}
\end{center}
\end{table}

\begin{figure*}[tb]
\begin{center}
\includegraphics[scale=0.5]{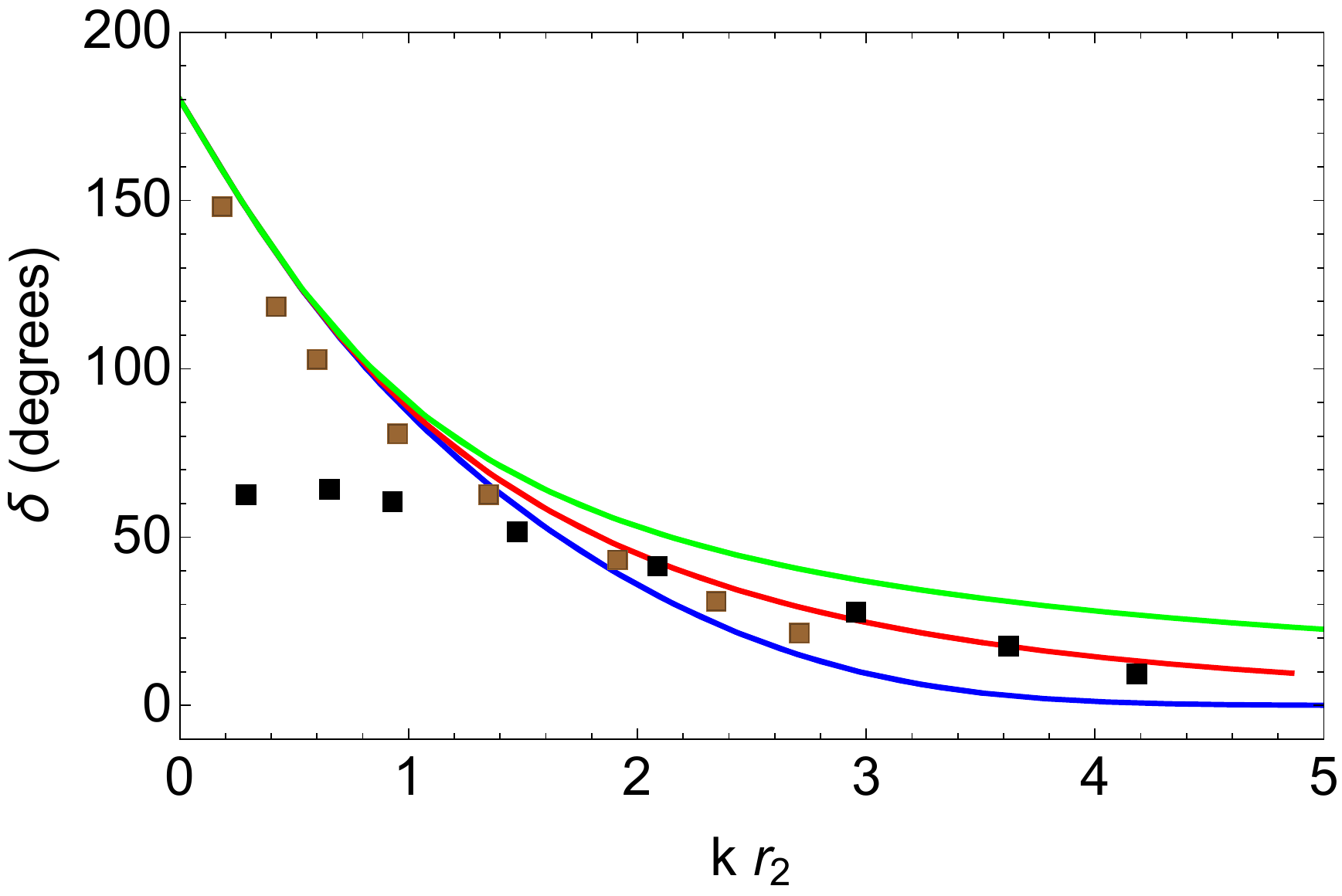}  
\end{center}
\caption{\label{fig1dim}
Phase shifts (in degrees) as a function of the center-of-mass momentum in units of the inverse effective range for square-root (green line), Yamaguchi (red line), and Gaussian (blue line) potentials. Physical values from the Granada phase-shift analysis \cite{NavarroPerez:2013usk}
for two nucleons in the $^3S_1$ (brown squares) and $^1S_0$ (black squares) channels are shown for comparison.
}
\end{figure*}

Similarly, for each of the potentials the phase shifts are universal as a function of the momentum $k$ in units of $1/r_2$. The phase shifts for the three potentials we consider are plotted in Fig. \ref{fig1dim}. The square-root potential gives the same phase shifts as the ERE truncated at the effective range. The Yamaguchi and Gaussian potentials give phase shifts that are essentially the same as those of the square-root potential for small momenta but decrease more rapidly as a consequence of increasing shape parameters. Despite the very different two-body pole structure, the dimensionless shape parameters are relatively small: $r_2^{3}/32$ for the Yamaguchi potential and $\simeq 0.05\, r_2^{3}$ for the Gaussian potential. 

The existence of a double pole would strictly require only the first two terms in the ERE, scattering length and effective range. This amounts to the square-root potential, Eq. \eqref{sqrtY}, as the LO in our expansion. We expect the shape parameter $P_2$ to be a subleading  effect. Since the typical momentum for the double pole is $\kappa_2 r_2\sim 1$ (from Eq. \eqref{B2} and Table \ref{beta2table}), a minimal condition for the shape-parameter term to be small compared to the first two terms is that $2P_2/r_2^3\ll 1$, suggesting a $\sim 10\%$ correction at the momenta of interest from the Yamaguchi and Gaussian potentials. Alternatively, $\kappa_2 r_2\simeq 0.8$ for these potentials, indicating instead an uncertainty of $\sim 20\%$ from deviations from the ERE truncated at the effective range. In any case, we can view these other potentials as incorporating some higher-order effects. In an EFT, this can be done without danger as long as no sensitivity to high-energy physics is introduced through regulator dependence, which is the case here. Few-body results might then not be very different among these potentials.

\section{Spinless three-body system}
\label{3spinless}

The EFT with a single two-body pole, characterized by a nonderivative two-body contact interaction at LO, produces three- and more-body observables with essential cutoff dependence, unless a nonderivative three-body contact interaction is also present at LO \cite{Bedaque:1998kg,Bedaque:1998km}. 
Incorporating the range at LO through a dimer field \cite{Kaplan:1996nv,Beane:2000fi}, no three-body force is needed for renormalization but the second pole leads to an unphysical three-body threshold that poses problems in the solution of the three-body equations \cite{Gabbiani:2001yh}. 
Here, the LO two-body amplitude not only has a sufficiently soft ultraviolet behavior but also does not suffer from an unphysical threshold because the redundant pole is not a solution of the Schr\"odinger equation. 
Since there is no longer a renormalization rationale to assume three- or more-body forces at LO, all energies 
should scale with the single two-body parameter controlling the $a_2=2r_2$ limit.

One of the first momentum-space calculations of the three-nucleon system was made by Sitenko and Kharchenko \cite{Sitenko:1963OnTB}. Employing Jacobi momenta
\begin{equation}
\vec{k}_{ij}=\frac{1}{2}\left(\vec{p}_i-\vec{p}_j\right),
\quad
\vec{k}_i=\frac{1}{3}\left(2\vec{p}_i-\vec{p}_j-\vec{p}_k\right),
\end{equation}
for $i\ne j\ne k$, where $\vec{p}_i$ stands for the momentum of particle $i$ in the center-of-mass frame of the three-body system, they solved the Faddeev equations \cite{Faddeev:1960su,Eyges:1961} in the case of a separable two-body potential. 
We follow the formulation of Ref. \cite{Sitenko:1963OnTB} closely --- see also the pedagogical review \cite{Timoteo:2021ika} --- to calculate the ground-state binding energy $B_3=\kappa_3/m$ when $a_2=2r_2$.

We start with the simplest case, that of three spinless particles. The wavefunction being symmetric under permutation of any pair of particles, it can be written as
\begin{equation}
    \psi_3 (\vec{p}_1,\vec{p}_2,\vec{p}_3)= 
    \psi(\vec{k}_{23},\vec{k}_1)
    +\psi(\vec{k}_{31},\vec{k}_2)
    +\psi(\vec{k}_{12},\vec{k}_3), 
    \quad 
    \psi(-\vec{k},\vec{k}_i)=\psi(\vec{k},\vec{k}_i).
\end{equation}
For a separable potential with only $S$-wave interactions, as we are interested in here, the wavefunction components are given by
\begin{equation}
\psi(p,q) = -\frac{\lambda g(p)}{\kappa_3^2 + p^2 + 3 q^2/4} \, a(q),
\end{equation}
where the profile function $a(q)$ is governed by a one-dimensional integral equation
\begin{equation}
a(q) = \frac{2}{\pi} \int_0^\infty dq' \, q'^2 \, {\cal K}(q,q';\kappa_3)\,  a(q') , 
\label{aq}
\end{equation}
with the kernel
\begin{equation}
{\cal K}(q,q';\kappa_3) =  \left[\Lambda^{-1}\left(\sqrt{\kappa_3^2+\frac{3q^2}{4}}\right)-\lambda^{-1}
\right]^{-1} \, 
\int_{-1}^1 dy \, 
\frac{g(\pi_2)\, g(\pi_1)}{\kappa_3^2+q^2+{q'}^2+qq'y}, 
\label{ang}
\end{equation}
\begin{equation}
\pi_1 = \sqrt{q^2/4 + {q'}^2 + q q' y}, 
\quad
\pi_2 = \sqrt{q^2 + {q'}^2/4 + q q' y}.
\end{equation}

Thanks to the factors of $g(p)$ in the kernel, Eq. \eqref{aq} is ultraviolet convergent and, as anticipated, no three-body force is needed for renormalization at LO. We compute the angular integration 
in Eq. \eqref{ang} 
numerically. The integral equation \eqref{aq} is solved on a momentum grid (with points labeled by $i$), a non-trivial solution existing provided
\begin{equation}
\det{\left[
\delta_{ij}-\bar {\cal K}(q_i,q_j';\kappa_3) \right] } =  0, 
\end{equation}
where 
$\bar{\cal K}(q_i,q'_j;\kappa_3)= (2/\pi){\cal K}(q_i,q'_j;\kappa_3) \, q_j^2 \, \Delta q_{j,j+1}'$ 
is the weighted kernel. The problem becomes a search for the value of $\kappa_3$ that makes the determinant vanish. The profile function $a(q)$ is then obtained 
from the eigenvectors. 

For $a_2=2r_2$, we typically find more than one solution. The single potential parameter implies that we can write for the deepest three-body bound state, analogously to Eq. \eqref{B2},
\begin{equation}
\frac{B_3}{3}= \frac{\beta_3}{3mr_2^2} 
=\frac{2\beta_3}{3\beta_2} \frac{B_2}{2} 
\equiv \varsigma_3 \frac{B_2}{2},
\label{B3}
\end{equation}
where $\varsigma_3$ is another potential-dependent dimensionless parameter. The values for $\varsigma_3$ are given in Table \ref{xitable} for the three potentials considered in this work. Our value for the square-root potential is not inconsistent with the results at smaller $r_2/a_2$ shown in Ref. \cite{Benayoun:1973tp}. Surprisingly, it falls in-between the potentials with softer ultraviolet behavior (and nonzero higher ERE parameters). The numbers for the various potentials are the same within about 5\% despite the difference in potential shapes, indicating an approximate universality. The small effects of higher ERE parameters must be amenable to a distorted-wave perturbative treatment around the square-root potential.


\begin{table}[tb]
\begin{center}
\begin{tabular}{c|c|c|c}
  &Square-Root & \text{Yamaguchi} & \text{Gaussian}    
  \\ 
  \hline
  $\varsigma_3$ & 3.73 & 3.80 & 3.54
  \\ 
\end{tabular}
\caption{
\label{xitable}
Slope $\varsigma_3$ of the correlation \eqref{B3} between three- and two-body binding energies per spinless particle for the square-root, Yamaguchi and Gaussian separable potentials, when $a_2=2r_2$. 
}
\end{center}
\end{table}

\section{Three-nucleon system}
\label{3nucleons}

The wavefunction for the three-nucleon system is antisymmetric under permutations of any of the three pairs of nucleons. Both total spin $S$ and total isospin $T$ may be either $1/2$ or $3/2$. 
Here we concentrate on the triton, which consists of one proton and two neutrons, with $T=1/2$ and $S=1/2$. 
In this case, there are contributions from two channels, a nucleon pair being in a spin-singlet state $s=0$ or in the triplet state $s=1$.
We take each channel to be governed by an $S$-wave separable potential of the type discussed in Sec. \ref{2bsys} but allow for different parameters in each two-nucleon channel, which we label with a subscript $0$ or $1$ according to its spin. Again, we expect ultraviolet convergence without three-nucleon forces, in contrast to the situation in Pionless EFT with standard power counting \cite{Bedaque:1999ve}. 
The wavefunction can then be written in terms of two components \cite{Sitenko:1963OnTB}
\begin{equation}
\left(
\begin{array}{c}
\psi_1\\
\psi_0
\end{array}
\right)
= -\frac{1}{\kappa_3^2 + p^2 + 3q^2/4} ~
\left(
\begin{array}{c}
\lambda_1\, g_1(p)\, a(q)\\
\lambda_0\, g_0(p)\, b(q)
\end{array}
\right).
\end{equation}
The two profile functions obey a set of coupled integral equations, 
\begin{equation}
\left(
\begin{array}{c}
a(q)\\
b(q)
\end{array}
\right)
= \frac{1}{2\pi} \int_0^\infty dq'{q'}^2 
\left(
\begin{array}{cc}
{\cal K}_{11}(q,q';\kappa_3) & 3{\cal K}_{10}(q,q';\kappa_3) \\
3{\cal K}_{01}(q,q';\kappa_3) & {\cal K}_{00}(q,q';\kappa_3) 
\end{array}
\right)
\left(
\begin{array}{c}
a(q')\\
b(q')
\end{array}
\right),
\label{coup}
\end{equation}
where
\begin{equation}
{\cal K}_{ss'}(q,q';\kappa_3) = 
\left[\Lambda_{s}^{-1}\left(\sqrt{\kappa_3^2+3q^2/4}\right)
-\lambda_{s}^{-1}\right]^{-1}
\int_{-1}^1 dy \, 
\frac{g_{s}(\pi_2)\,g_{s'}(\pi_1)}{\kappa_3^2+p^2+{q'}^2+qq'y}. 
\label{angspins}
\end{equation}

Again, the system described in Eq. \eqref{coup} has a non-trivial solution provided
\begin{equation}
\det 
\left(
\begin{array}{cc}
{\bf 1}-\bar{\cal K}_{11}(q,q';\kappa_3) & -3\bar{\cal K}_{10}(q,q';\kappa_3)
\\
-3\bar{\cal K}_{01}(q,q';\kappa_3) & {\bf 1}-\bar{\cal K}_{00}(q,q';\kappa_3)
\end{array}
\right) =  0, 
\end{equation}
where 
$\bar{\cal K}_{ss'}(q,q';\kappa_3)= (1/2\pi){\cal K}_{ss'}(q,q';\kappa_3) \, q'^2 dq'$, and the profile functions are obtained from the eigenvectors of the block matrix. 
The value of 
$\kappa_3$ for which the determinant is zero corresponds to the triton binding energy $B=\kappa_3^2/m_N$, where $m_N$ is the nucleon mass.

In the following, we consider the case when there are two independent dimensionful parameters because $a_{2s}=2r_{2s}$ for both $s=0,1$. We typically find various bound states at the same values of the ratio $r_{21}/r_{20}$. We postpone a more comprehensive analysis of the structure of these states to a future publication. In the following, we consider two cases of interest to nuclear physics.
We compare our results for the three-nucleon system with those obtained in LQCD at an unphysical pion mass $m_\pi \simeq 806$ MeV and with experiment at physical pion mass.

\section{Lattice QCD predictions for two and three nucleons}
\label{unphysical}

At the quark masses for which $m_\pi \simeq 806$ MeV, the nucleon mass is
$m_N=1.634(0)(0)(18)$ GeV, with errors corresponding to statistics, fitting systematics, and lattice spacing \cite{NPLQCD:2012mex}. The two-nucleon ERE parameters are found \cite{NPLQCD:2013bqy} to be somewhat large on the scale set by the pion Compton wavelength $m_\pi^{-1}\simeq 0.25$ fm:
\begin{eqnarray}
&&a_{21} = 1.82^{+0.14+0.17}_{-0.13-0.12} ~{\rm fm},
\quad
r_{21} = 0.906^{+0.068+0.068}_{-0.075-0.084} ~{\rm fm},
\label{LQCDEREparam1}
\\
&&a_{20} = 2.33^{+0.19+0.27}_{-0.17-0.20} ~{\rm fm},
\quad
r_{20} = 1.130^{+0.071+0.059}_{-0.077-0.063} ~{\rm fm},
\label{LQCDEREparam}
\end{eqnarray}
suggesting that a Pionless EFT with power counting based on $r_{2s}={\cal O}(\aleph^{-1})$ should hold. 
(The results from Ref. \cite{Wagman:2017tmp} are consistent within errors.) 

The ERE parameters satisfy $a_{2s}=2r_{2s}>0$ within errors. To the extent that higher ERE parameters are small \cite{NPLQCD:2013bqy}, we expect that for each two-nucleon $S$-wave channel there are two near-degenerate poles in the positive imaginary axis, approximately described by the square-root potential. The two-nucleon binding energies obtained from these ERE parameter values are given in Table \ref{tab1} for the three separable potentials we consider here. There is a $\approx 25\%$ spread in predictions, which fall into the errors of the direct LQCD determination \cite{NPLQCD:2012mex}, also shown in the table. Table \ref{shape} gives the values of the corresponding shape parameters, which lie well inside the range of values from Ref. \cite{NPLQCD:2013bqy}. For $a_{2s}=2r_{2s}$, the phase shifts in the two channels can be read off Fig. \ref{fig1dim} when $r_2$ is replaced by the respective effective-range value. The phase shifts do not turn negative as suggested by the higher-energy LQCD data.
Nevertheless, the closeness of the predictions from the various potentials to each other and to direct LQCD values at low and moderate energies suggest that the latter can be reproduced in perturbation theory around the LO square-root potential. 

\begin{table}[bt]
\begin{center}
\begin{tabular}{|c|c|c|c|c|c|}
\hline
   & Square-Root & Yamaguchi & Gaussian  & LQCD
  \\ 
  \hline
  $B_{21}$/MeV & 25.3 & 19.5 &  18.4  &  19.5~(3.6)~(3.1)~(0.2)
  \\ 
  \hline
  $B_{20}$/MeV & 12.7  & 11.1 & 10.7 & 15.9~(2.7)~(2.7)~(0.2)
  \\ 
  \hline
\end{tabular}
\caption{\label{tab1}
Two-nucleon binding energies $B_{2s}$ (in MeV) for the spin $s=0,1$ channels supported by the square-root, Yamaguchi, and Gaussian separable potentials with ERE parameters from lattice QCD \cite{NPLQCD:2013bqy}, compared to the direct LQCD value \cite{NPLQCD:2012mex}.
}
\end{center}
\end{table}


\begin{table}[bt]
\begin{center}
\begin{tabular}{|c|c|c|c|c|c|}
\hline
    & Yamaguchi & Gaussian  & LQCD
  \\ 
  \hline
  $P_{1}~({\rm fm}^3)$ & 0.023
  &  0.037
  &  [$-$0.147,
  0.176]
  \\ 
  \hline
  $P_{0}~({\rm fm}^3)$ & 0.044
  & 0.070
  & [$-$0.205,
  0.117]
  \\ 
  \hline
\end{tabular}
\caption{\label{shape}
Shape parameters $P_{s}$ (in ${\rm fm}^3$) in spin $s=0,1$ channels for the Yamaguchi and Gaussian separable potentials with ERE parameters from lattice QCD \cite{NPLQCD:2013bqy}, compared to the range of LQCD values \cite{NPLQCD:2013bqy}.
}
\end{center}
\end{table}

Table \ref{tab2} shows results for the binding energy of the deepest three-body bound state we found with the Gaussian separable potential, where we varied the ERE parameters \eqref{LQCDEREparam} within error bars (errors added in quadrature). The relatively large uncertainties in the effective ranges generate large uncertainties in binding energies. The Gaussian potential yields the most stable results, but other potentials produce similar outcomes. In Table \ref{tab3}, central results for the three potentials are compared with the direct LQCD value \cite{NPLQCD:2012mex}. Although uncertainties are large, central values for the three potentials are extremely close, and well within the range of values obtained directly from LQCD \cite{NPLQCD:2012mex}. These results add further support to the view that the NPLQCD results at these unphysical quark masses can be described by a nonlocal potential with a near-degenerate double two-body pole.

\begin{table}[bt]
\begin{center}
\begin{tabular}{|c|c|c|c|}
\hline
 $3S1~\symbol{92}~1S0$ & \text{Lower} & \text{Central} & \text{Upper}    \\ \hline
 \text{Lower} & 26.1   
 & 57.4 
 & 52.0 
 \\
 \text{Central} & 67.3   
 & 56.5 
 & 89.8 
 \\
 \text{Upper} & 59.2 
 & 57.1 
 & 52.6 
 \\ \hline
\end{tabular}
\caption{\label{tab2}
Three-nucleon binding energy (in MeV) supported by the Gaussian separable potential  with ERE parameters from lattice QCD values \cite{NPLQCD:2013bqy}. ``Lower'', ``central'', and ``upper'' refer to the minimum, central, and maximum values of the scattering lengths $a_{2s}$ and effective ranges $r_{2s}$ (in $^3S_1$ and $^1S_0$ channels) .}
\end{center}
\end{table}

\begin{table}[bt]
\begin{center}
\begin{tabular}{|c|c|c|c|c|c|}
\hline
  & Square-Root & Yamaguchi & Gaussian  & LQCD
  \\ 
  \hline $B_3$/MeV 
  &     56.5        &    56.6      &   56.5    &  53.9~(7.1)~(8.0)~(0.6)
  \\ 
  \hline
\end{tabular}
\caption{\label{tab3}
Three-nucleon binding energy $B_3$ (in MeV) supported by the square-root, Yamaguchi, and Gaussian separable potentials with ERE parameters from lattice QCD values \cite{NPLQCD:2013bqy}, compared to the direct LQCD value \cite{NPLQCD:2012mex}.
}
\end{center}
\end{table}

\section{Physical pion mass}
\label{physical}

At physical quark masses, the average nucleon mass is 
$m_N\simeq 938.92$ MeV \cite{ParticleDataGroup:2020ssz}, while the two-nucleon ERE parameters are extracted from data as \cite{deSwart:1995ui,Babenko:2010}
\begin{eqnarray}
&& a_{21} = 5.4194(20) ~{\rm fm},
\quad
r_{21} = 1.7536(25) ~{\rm fm},
\label{exptEREparam1}
\\
&& a_{20} = -23.7154(80) ~{\rm fm},
\quad
r_{20} = 2.706(67) ~{\rm fm}.
\label{exptEREparam}
\end{eqnarray}
While the scattering lengths are certainly large compared to 
$m_\pi^{-1}\simeq 1.4$ fm, the situation is less clear cut for the effective ranges. It has been suggested \cite{Kaplan:1996nv,Beane:2000fi} that an expansion based on $r_{2s}={\cal O}(\aleph^{-1})$ might be more effective  than the standard Pionless EFT expansion with $r_{2s}={\cal O}(R)$. However, its implementation through energy-dependent interactions is not optimal for many-body applications. The formulation through a separable potential considered here and in Ref. \cite{Beane:2021dab} might be useful. Analogous statements can be made about Chiral EFT for energy- \cite{Beane:2001bc,Long:2013cya} and momentum- \cite{SanchezSanchez:2020kbx,Yang:2020pgi,Yang:2021vxa,Peng:2021pvo} dependent interactions.  

The triton binding energy with input similar to Eqs. \eqref{exptEREparam1} and \eqref{exptEREparam} has been discussed many times in the literature, for example Ref. \cite{Bander:1964gb} for the square-root potential, Ref. \cite{Sitenko:1963OnTB} for Yamaguchi, and Ref. \cite{Timoteo:2021ika} for Gaussian. Results fall in a wide range $\simeq 7-12$ MeV, a sensitivity to the two-nucleon interaction that is encapsulated in the so-called Phillips line \cite{Phillips:1968zze}. We therefore focus here on the possibility of an expansion around $a_{2s}=2r_{2s}$. The $^3S_1$ channel is not far from this limit. In $^1S_0$, one finds a very shallow pole at negative imaginary momentum, while $a_{20}=2r_{20}$ gives relatively shallow poles at positive imaginary momentum thanks to the relatively large effective range. One might hope that these departures from two-body data at very low energies are not so important for the deeper ground states of larger nuclei. It is known that for the Yamaguchi potential there is little sensitivity of the triton binding energy to $a_{20}$, at least when the latter is large and negative \cite{Kharchenko:1968}.

In Table \ref{tab4} we give the two-body binding energies for our separable potentials when $r_{2s}$ is fixed by the empirical values in Eqs. \eqref{exptEREparam1} and \eqref{exptEREparam}, and $a_{2s}=2r_{2s}$. We also compare our results with the experimental binding energy of the deuteron \cite{deSwart:1995ui}. As one would expect from the shallowness of the observed deuteron and $^1S_0$ virtual state, our unrealistic scattering lengths lead to deeper bound states in both cases. Clearly the very low-energy region cannot be described well by our potentials. The issue is whether they capture the physics at higher energies and deeper bound states as the triton. Indeed, for the three potentials we find a state with energy close to the triton's \cite{Purcell:2010}, as shown in Table \ref{tab5}. 
In contrast to nature, however, this triton is unstable to decay into the $^3S_1$ two-nucleon state, as it has a smaller binding energy for all three potentials.
The reason for the instability is, of course, the unrealistic two-body binding energies. It is remarkable that the triton energy is much less sensitive to the potential, varying by at most 20\%. 
For Yamaguchi and Gaussian potentials, the result is particularly close to experiment and the triton is unstable by less than 5\% of the total energy. It is not inconceivable that higher-order terms will change the $^3S_1$ binding energy sufficiently to bring it below the triton's. Of course, even if that is the case, one would have to investigate to which extent this EFT would still apply at the low energies relevant to the observed two-nucleon states. 

\begin{table}[bt]
\begin{center}
\begin{tabular}{|c|c|c|c|c|c|}
\hline
   & Square-Root & Yamaguchi & Gaussian  & experiment
  \\ 
  \hline
  $B_{21}$/MeV & 13.4949 & 9.26145 &  8.69714 &  2.224575(9)
  \\ 
  \hline
  $B_{20}$/MeV & 5.66342  & 3.88675 & 3.64993 &  ---
  \\ 
  \hline
\end{tabular}
\caption{\label{tab4}
Two-nucleon binding energies $B_{2s}$ (in MeV) for the spin $s=0,1$ channels, supported by the square-root, Yamaguchi, and Gaussian separable potentials with empirical $r_{2s}$ parameters \cite{deSwart:1995ui,Babenko:2010} and $a_{2s}=2r_{2s}$, compared to the experimental value \cite{deSwart:1995ui}.
}
\end{center}
\end{table}

The agreement with the empirical triton energy is likely a manifestation of the folklore that nuclear ground states are not very sensitive to two-nucleon scattering near threshold, which in conventional Pionless EFT is incorporated through an expansion around the unitarity limit \cite{Konig:2016utl,Konig:2019xxk}.
The phase shifts for $a_{2s}=2r_{2s}$ can again be read off Fig. \ref{fig1dim} once the physical values for $r_{2s}$ are used. To facilitate the comparison with empirical values, we plot in Fig. \ref{fig1dim} the results of the Granada phase-shift analysis \cite{NavarroPerez:2013usk} for the $^3S_1$ and $^1S_0$ channels. We see that all three potentials are close to the empirical $^3S_1$ phase shifts despite our scattering length not taking the experimental value. The three potentials also give $^1S_0$ phase shifts close to empirical for $kr_{20}\simge 1$, differences at smaller momenta reflecting the existence of a bound state when $a_{20}=2r_{20}>0$. The fact that these small-momentum differences do not dramatically affect our results for the triton binding energy supports existing folklore. Still, it is surprising that such a good agreement is found in the same framework that accommodates LQCD results.

\begin{table}[tb]
\begin{center}
\begin{tabular}{|c|c|c|c|c|c|}
\hline
  & Square-Root & Yamaguchi & Gaussian  & experiment
  \\ 
  \hline
  $B_3$/MeV & 7.496939 & 8.945608 & 8.397675 &  8.481798(2)
  \\ 
  \hline
\end{tabular}
\caption{\label{tab5}
Triton binding energy $B_3$ (in MeV) supported by the square-root, Yamaguchi, and Gaussian separable potentials with empirical $r_{2s}$ parameters \cite{deSwart:1995ui,Babenko:2010} and $a_{2s}=2r_{2s}$, compared to experiment \cite{Purcell:2010}.
}
\end{center}
\end{table}

\section{Conclusion} 
\label{conc}

Motivated by the NPLQCD data \cite{NPLQCD:2012mex,NPLQCD:2013bqy} at unphysical quark masses, we developed here a reorganization of Pionless EFT based on the presence of a nearly degenerate pair of two-body $S$-matrix poles on the positive imaginary axis of the complex momentum plane. The limit we consider, $a_2/r_2=2$, constitutes a different class of universality than the two-body unitarity limit $a_2/r_2\to \infty$. For the latter, the binding energy $B_A$ of an $A$-boson system scales with the three-body binding energy $B_3$: $3B_A/AB_3 =\xi_A$, with $\xi_A$ universal, dimensionless numbers \cite{vonStecher:2010,Carlson:2017txq}. Here, instead, $B_A$ scales with $B_2$ according to other universal numbers: $2B_A/AB_2 =\varsigma_A$. We have obtained $\varsigma_3\simeq 3.7$ for the deepest bound state, which is for no known reason close to $\xi_4\simeq 3.5$. Similar calculations could be performed for larger systems to obtain $\varsigma_{A\ge 4}$.

For systems with more $S$-wave channels, the dimensionless numbers $\varsigma_A$ become functions of the ratios of effective ranges. We find several bound states in the nuclear case of two channels. The NPLQCD effective-range parameters \cite{NPLQCD:2013bqy} are close to the limit we consider here, with $r_{21}/r_{20}\simeq 0.8$. 
Although LQCD errors are large, our binding energies for the two-body and deepest three-body states, based solely on two-nucleon scattering input, are in good agreement with the corresponding, direct NPLQCD results. To the extend that strangeness channels also display $a_2/r_2\approx 2$ \cite{Wagman:2017tmp}, a similar calculation could be carried out for three-body hypernuclei. 

Of course, this consistency does not, by itself, dispel worries \cite{Nicholson:2021zwi} about the validity of the lattice simulations we used as input. It has been argued \cite{Iritani:2017rlk} that most LQCD two-nucleon results are inconsistent because they contain $S$-matrix poles with the wrong residue. This criticism does not apply to the double pole of NPLQCD \cite{Wagman:2017tmp,Beane:2017edf}, and in any case it implicitly assumes the potential to be local (in the sense of containing a finite number of derivatives) and decreasing sufficiently fast at large distances (cf. Ref. \cite{Bargmann:1949}). As we have made explicit above, there is no inconsistency for a nonlocal potential, which with two parameters per channel can also accommodate other LQCD results, where the two-body poles are separated
\cite{Berkowitz:2015eaa,Horz:2020zvv}. The consistency we found here between two- and three-body NPLQCD results suggests that whatever mechanism is responsible for the unusual two-nucleon scattering parameters is also responsible for two- and three-nucleon binding energies. It does not seem likely that the NPLQCD results arise simply from wrong plateau identifications. The nonlocality of the potentials we consider could explain why the same poles are not found when LQCD simulations are interpreted in terms of a local potential \cite{Iritani:2017rlk}. 

Although for physical quark masses two-nucleon scattering lengths and effective ranges do not satisfy $a_{2s}=2r_{2s}$, the fine-tuning in $a_{2s}$ could be less important in few-body systems than for the deuteron. To test this idea, we also calculated the triton binding energy with empirical values for $r_{2s}$ and unphysical $a_{2s}=2r_{2s}$. This case is essentially the same as NPLQCD's, except that $r_{21}/r_{20}\simeq 0.65$. Surprisingly, for any of the potentials we considered we found a bound state with energy close to the triton's, a result that probably reflects the reasonable description of phase shifts around $kr_{2s}/2\sim 1$ seen in Fig. \ref{fig1dim}. 
This bound state is unstable because of the unrealistically large $^3S_1$ two-body binding energy. Since its energy varies much less with the potential than the two-body energies, there is hope the latter can be corrected sufficiently at higher orders.

These tantalizing outcomes suggest several additional investigations. Even at the leading order we examined here, it would be interesting to map out the universal features of excited three-body levels. For both unphysical and physical pion masses, the similarities among the various 
potentials, which differ only by small higher effective-range expansion parameters, suggest that a distorted-wave perturbation around the square-root potential \cite{Beane:2021dab} might be successful. The extension to higher-body binding energies would shed light on the significance of the agreement with data we have found so far. And, if our results turn out not to be accidental, one would have to understand the origin of this surprising non-locality.

%
%

\section*{Acknowledgements}
UvK thanks Silas Beane, Sebastian K\"onig, and Bingwei Long for useful discussions.
We are thankful to Wael Elkamhawy for pointing out an error in a previous version of the manuscript.
VST is grateful to Prof. Cristiane Smith for hospitality
at the Universiteit Utrecht and to UvK for hospitality at the Institut de Physique Nucleaire Orsay, where parts of this work were carried out.
This work is supported in part by 
FAPESP under grant number 2019/10889-1 (VST), 
CNPq under grant number 305004/2022-0 (VST),
and the U.S. Department of Energy, Office of Science, Office of Nuclear Physics under award number DE-FG02-04ER41338 (UvK).

\end{document}